
\vsize=8.75truein
%
\catcode`@=11 
%
%
%

\font\fourteenrm=cmr10 scaled\magstep2
\font\twelverm=cmr10 scaled\magstep1
\font\ninerm=cmr9            \font\sixrm=cmr6

\font\fourteenbf=cmbx10 scaled\magstep2
\font\twelvebf=cmbx10 scaled\magstep1
\font\ninebf=cmbx9            \font\sixbf=cmbx6
\font\seventeeni=cmmi10 scaled\magstep3     \skewchar\seventeeni='177
\font\fourteeni=cmmi10 scaled\magstep2      \skewchar\fourteeni='177
\font\twelvei=cmmi10 scaled\magstep1        \skewchar\twelvei='177
\font\ninei=cmmi9                           \skewchar\ninei='177
\font\sixi=cmmi6                            \skewchar\sixi='177
\font\seventeensy=cmsy10 scaled\magstep3    \skewchar\seventeensy='60
\font\fourteensy=cmsy10 scaled\magstep2     \skewchar\fourteensy='60
\font\twelvesy=cmsy10 scaled\magstep1       \skewchar\twelvesy='60
\font\ninesy=cmsy9                          \skewchar\ninesy='60
\font\sixsy=cmsy6                           \skewchar\sixsy='60

\font\fourteenex=cmex10 scaled\magstep2
\font\twelveex=cmex10 scaled\magstep1

\font\fourteensl=cmsl10 scaled\magstep2
\font\twelvesl=cmsl10 scaled\magstep1
\font\ninesl=cmsl9

\font\fourteenit=cmti10 scaled\magstep2
\font\twelveit=cmti10 scaled\magstep1
\font\twelvett=cmtt10 scaled\magstep1
\font\twelvecp=cmcsc10 scaled\magstep1
\font\tencp=cmcsc10
\newfam\cpfam
%
%
\newcount\f@ntkey            \f@ntkey=0
\def\samef@nt{\relax \ifcase\f@ntkey \rm \or\oldstyle \or\or
         \or\it \or\sl \or\bf \or\tt \or\caps \fi }
\def\fourteenpoint{\relax
    \textfont0=\fourteenrm          \scriptfont0=\tenrm
    \scriptscriptfont0=\sevenrm
     \def\rm{\fam0 \fourteenrm \f@ntkey=0 }\relax
    \textfont1=\fourteeni           \scriptfont1=\teni
    \scriptscriptfont1=\seveni
     \def\oldstyle{\fam1 \fourteeni\f@ntkey=1 }\relax
    \textfont2=\fourteensy          \scriptfont2=\tensy
    \scriptscriptfont2=\sevensy
    \textfont3=\fourteenex     \scriptfont3=\fourteenex
    \scriptscriptfont3=\fourteenex
    \def\it{\fam\itfam \fourteenit\f@ntkey=4 }\textfont\itfam=\fourteenit
    \def\sl{\fam\slfam \fourteensl\f@ntkey=5 }\textfont\slfam=\fourteensl
    \scriptfont\slfam=\tensl
    \def\bf{\fam\bffam \fourteenbf\f@ntkey=6 }\textfont\bffam=\fourteenbf
    \scriptfont\bffam=\tenbf     \scriptscriptfont\bffam=\sevenbf
    \def\tt{\fam\ttfam \twelvett \f@ntkey=7 }\textfont\ttfam=\twelvett
    \h@big=11.9\p@{} \h@Big=16.1\p@{} \h@bigg=20.3\p@{} \h@Bigg=24.5\p@{}
    \def\caps{\fam\cpfam \twelvecp \f@ntkey=8 }\textfont\cpfam=\twelvecp
    \setbox\strutbox=\hbox{\vrule height 12pt depth 5pt width\z@}
    \samef@nt}
\def\twelvepoint{\relax
    \textfont0=\twelverm          \scriptfont0=\ninerm
    \scriptscriptfont0=\sixrm
     \def\rm{\fam0 \twelverm \f@ntkey=0 }\relax
    \textfont1=\twelvei           \scriptfont1=\ninei
    \scriptscriptfont1=\sixi
     \def\oldstyle{\fam1 \twelvei\f@ntkey=1 }\relax
    \textfont2=\twelvesy          \scriptfont2=\ninesy
    \scriptscriptfont2=\sixsy
    \textfont3=\twelveex          \scriptfont3=\twelveex
    \scriptscriptfont3=\twelveex
    \def\it{\fam\itfam \twelveit \f@ntkey=4 }\textfont\itfam=\twelveit
    \def\sl{\fam\slfam \twelvesl \f@ntkey=5 }\textfont\slfam=\twelvesl
    \scriptfont\slfam=\ninesl
    \def\bf{\fam\bffam \twelvebf \f@ntkey=6 }\textfont\bffam=\twelvebf
    \scriptfont\bffam=\ninebf     \scriptscriptfont\bffam=\sixbf
    \def\tt{\fam\ttfam \twelvett \f@ntkey=7 }\textfont\ttfam=\twelvett
    \h@big=10.2\p@{}
    \h@Big=13.8\p@{}
    \h@bigg=17.4\p@{}
    \h@Bigg=21.0\p@{}
    \def\caps{\fam\cpfam \twelvecp \f@ntkey=8 }\textfont\cpfam=\twelvecp
    \setbox\strutbox=\hbox{\vrule height 10pt depth 4pt width\z@}
    \samef@nt}
\def\tenpoint{\relax
    \textfont0=\tenrm          \scriptfont0=\sevenrm
    \scriptscriptfont0=\fiverm
    \def\rm{\fam0 \tenrm \f@ntkey=0 }\relax
    \textfont1=\teni           \scriptfont1=\seveni
    \scriptscriptfont1=\fivei
    \def\oldstyle{\fam1 \teni \f@ntkey=1 }\relax
    \textfont2=\tensy          \scriptfont2=\sevensy
    \scriptscriptfont2=\fivesy
    \textfont3=\tenex          \scriptfont3=\tenex
    \scriptscriptfont3=\tenex
    \def\it{\fam\itfam \tenit \f@ntkey=4 }\textfont\itfam=\tenit
    \def\sl{\fam\slfam \tensl \f@ntkey=5 }\textfont\slfam=\tensl
    \def\bf{\fam\bffam \tenbf \f@ntkey=6 }\textfont\bffam=\tenbf
    \scriptfont\bffam=\sevenbf     \scriptscriptfont\bffam=\fivebf
    \def\tt{\fam\ttfam \tentt \f@ntkey=7 }\textfont\ttfam=\tentt
    \def\caps{\fam\cpfam \tencp \f@ntkey=8 }\textfont\cpfam=\tencp
    \setbox\strutbox=\hbox{\vrule height 8.5pt depth 3.5pt width\z@}
    \samef@nt}
%
%
%
%
\newdimen\h@big  \h@big=8.5\p@
\newdimen\h@Big  \h@Big=11.5\p@
\newdimen\h@bigg  \h@bigg=14.5\p@
\newdimen\h@Bigg  \h@Bigg=17.5\p@
\def\big#1{{\hbox{$\left#1\vbox to\h@big{}\right.\n@space$}}}
\def\Big#1{{\hbox{$\left#1\vbox to\h@Big{}\right.\n@space$}}}
\def\bigg#1{{\hbox{$\left#1\vbox to\h@bigg{}\right.\n@space$}}}
\def\Bigg#1{{\hbox{$\left#1\vbox to\h@Bigg{}\right.\n@space$}}}
%
%
%
\normalbaselineskip = 20pt plus 0.2pt minus 0.1pt
\normallineskip = 1.5pt plus 0.1pt minus 0.1pt
\normallineskiplimit = 1.5pt
\newskip\normaldisplayskip
\normaldisplayskip = 20pt plus 5pt minus 10pt
\newskip\normaldispshortskip
\normaldispshortskip = 6pt plus 5pt
\newskip\normalparskip
\normalparskip = 6pt plus 2pt minus 1pt
\newskip\skipregister
\skipregister = 5pt plus 2pt minus 1.5pt
\newif\ifsingl@    \newif\ifdoubl@
\newif\iftwelv@    \twelv@true
\def\singlespace{\singl@true\doubl@false\spaces@t}
\def\doublespace{\singl@false\doubl@true\spaces@t}
\def\normalspace{\singl@false\doubl@false\spaces@t}
\def\Tenpoint{\tenpoint\twelv@false\spaces@t}
\def\Twelvepoint{\twelvepoint\twelv@true\spaces@t}
\def\spaces@t{\relax%
 \iftwelv@ \ifsingl@\subspaces@t3:4;\else\subspaces@t1:1;\fi%
 \else \ifsingl@\subspaces@t3:5;\else\subspaces@t4:5;\fi \fi%
 \ifdoubl@ \multiply\baselineskip by 5%
 \divide\baselineskip by 4 \fi \unskip}
\def\subspaces@t#1:#2;{
      \baselineskip = \normalbaselineskip
      \multiply\baselineskip by #1 \divide\baselineskip by #2
      \lineskip = \normallineskip
      \multiply\lineskip by #1 \divide\lineskip by #2
      \lineskiplimit = \normallineskiplimit
      \multiply\lineskiplimit by #1 \divide\lineskiplimit by #2
      \parskip = \normalparskip
      \multiply\parskip by #1 \divide\parskip by #2
      \abovedisplayskip = \normaldisplayskip
      \multiply\abovedisplayskip by #1 \divide\abovedisplayskip by #2
      \belowdisplayskip = \abovedisplayskip
      \abovedisplayshortskip = \normaldispshortskip
      \multiply\abovedisplayshortskip by #1
        \divide\abovedisplayshortskip by #2
      \belowdisplayshortskip = \abovedisplayshortskip
      \advance\belowdisplayshortskip by \belowdisplayskip
      \divide\belowdisplayshortskip by 2
      \smallskipamount = \skipregister
      \multiply\smallskipamount by #1 \divide\smallskipamount by #2
      \medskipamount = \smallskipamount \multiply\medskipamount by 2
      \bigskipamount = \smallskipamount \multiply\bigskipamount by 4 }
\def\normalbaselines{ \baselineskip=\normalbaselineskip
   \lineskip=\normallineskip \lineskiplimit=\normallineskip
   \iftwelv@\else \multiply\baselineskip by 4 \divide\baselineskip by 5
     \multiply\lineskiplimit by 4 \divide\lineskiplimit by 5
     \multiply\lineskip by 4 \divide\lineskip by 5 \fi }
\Twelvepoint  
\interlinepenalty=50
\interfootnotelinepenalty=5000
\predisplaypenalty=9000
\postdisplaypenalty=500
\hfuzz=1pt
\vfuzz=0.2pt
%
%
%
\def\pagecontents{
   \ifvoid\topins\else\unvbox\topins\vskip\skip\topins\fi
   \dimen@ = \dp255 \unvbox255
   \ifvoid\footins\else\vskip\skip\footins\footrule\unvbox\footins\fi
   \ifr@ggedbottom \kern-\dimen@ \vfil \fi }
\def\makeheadline{\vbox to 0pt{ \skip@=\topskip
      \advance\skip@ by -12pt \advance\skip@ by -2\normalbaselineskip
      \vskip\skip@ \line{\vbox to 12pt{}\the\headline} \vss
      }\nointerlineskip}
\def\makefootline{\baselineskip = 1.5\normalbaselineskip
                 \line{\the\footline}}
\newif\iffrontpage
\newif\ifletterstyle
\newif\ifp@genum
\def\nopagenumbers{\p@genumfalse}
\def\pagenumbers{\p@genumtrue}
\pagenumbers
\newtoks\paperheadline
\newtoks\letterheadline
\newtoks\letterfrontheadline
\newtoks\lettermainheadline
\newtoks\paperfootline
\newtoks\letterfootline
\newtoks\date
\footline={\ifletterstyle\the\letterfootline\else\the\paperfootline\fi}
\paperfootline={\hss\iffrontpage\else\ifp@genum\tenrm\folio\hss\fi\fi}
\letterfootline={\hfil}
\headline={\ifletterstyle\the\letterheadline\else\the\paperheadline\fi}
\paperheadline={\hfil}
\letterheadline{\iffrontpage\the\letterfrontheadline
     \else\the\lettermainheadline\fi}
\lettermainheadline={\rm\ifp@genum page \ \folio\fi\hfil\the\date}
\def\monthname{\relax\ifcase\month 0/\or January\or February\or
   March\or April\or May\or June\or July\or August\or September\or
   October\or November\or December\else\number\month/\fi}
\date={\monthname\ \number\day, \number\year}
\countdef\pagenumber=1  \pagenumber=1
\def\advancepageno{\global\advance\pageno by 1
   \ifnum\pagenumber<0 \global\advance\pagenumber by -1
    \else\global\advance\pagenumber by 1 \fi \global\frontpagefalse }
\def\folio{\ifnum\pagenumber<0 \romannumeral-\pagenumber
           \else \number\pagenumber \fi }
\def\footrule{\dimen@=\prevdepth\nointerlineskip
   \vbox to 0pt{\vskip -0.25\baselineskip \hrule width 0.35\hsize \vss}
   \prevdepth=\dimen@ }
\newtoks\foottokens
\foottokens={\Tenpoint\singlespace}
\newdimen\footindent
\footindent=24pt
\def\vfootnote#1{\insert\footins\bgroup  \the\foottokens
   \interlinepenalty=\interfootnotelinepenalty \floatingpenalty=20000
   \splittopskip=\ht\strutbox \boxmaxdepth=\dp\strutbox
   \leftskip=\footindent \rightskip=\z@skip
   \parindent=0.5\footindent \parfillskip=0pt plus 1fil
   \spaceskip=\z@skip \xspaceskip=\z@skip
   \Textindent{$ #1 $}\footstrut\futurelet\next\fo@t}
\def\Textindent#1{\noindent\llap{#1\enspace}\ignorespaces}
\def\footnote#1{\attach{#1}\vfootnote{#1}}

\def\foot{\attach\footsymbolgen\vfootnote{\footsymbol}}
\let\footsymbol=\star
\newcount\lastf@@t           \lastf@@t=-1
\newcount\footsymbolcount    \footsymbolcount=0
\newif\ifPhysRev
\def\footsymbolgen{\relax \ifPhysRev \iffrontpage \NPsymbolgen\else
      \PRsymbolgen\fi \else \NPsymbolgen\fi
   \global\lastf@@t=\pageno \footsymbol }
\def\NPsymbolgen{\ifnum\footsymbolcount<0 \global\footsymbolcount=0\fi
   {\iffrontpage \else \advance\lastf@@t by 1 \fi
    \ifnum\lastf@@t<\pageno \global\footsymbolcount=0
     \else \global\advance\footsymbolcount by 1 \fi }
   \ifcase\footsymbolcount \fd@f\star\or \fd@f\dagger\or \fd@f\ast\or
    \fd@f\ddagger\or \fd@f\natural\or \fd@f\diamond\or \fd@f\bullet\or
    \fd@f\nabla\else \fd@f\dagger\global\footsymbolcount=0 \fi }
\def\fd@f#1{\xdef\footsymbol{#1}}
\def\PRsymbolgen{\ifnum\footsymbolcount>0 \global\footsymbolcount=0\fi
      \global\advance\footsymbolcount by -1
      \xdef\footsymbol{\sharp\number-\footsymbolcount} }
\def\space@ver#1{\let\@sf=\empty \ifmmode #1\else \ifhmode
   \edef\@sf{\spacefactor=\the\spacefactor}\unskip${}#1$\relax\fi\fi}
\def\attach#1{\space@ver{\strut^{\mkern 2mu #1} }\@sf\ }
%
%
%
\newcount\chapternumber      \chapternumber=0
\newcount\sectionnumber      \sectionnumber=0
\newcount\equanumber         \equanumber=0
\let\chapterlabel=0
\newtoks\chapterstyle        \chapterstyle={\Number}
\newskip\chapterskip         \chapterskip=\bigskipamount
\newskip\sectionskip         \sectionskip=\medskipamount
\newskip\headskip            \headskip=8pt plus 3pt minus 3pt
\newdimen\chapterminspace    \chapterminspace=15pc
\newdimen\sectionminspace    \sectionminspace=10pc
\newdimen\referenceminspace  \referenceminspace=25pc
\def\chapterreset{\global\advance\chapternumber by 1
   \ifnum\the\equanumber<0 \else\global\equanumber=0\fi
   \sectionnumber=0 \makel@bel}
\def\makel@bel{\xdef\chapterlabel{%
\the\chapterstyle{\the\chapternumber}.}}
\def\sectionlabel{\number\sectionnumber \quad }
\def\alphabetic#1{\count255='140 \advance\count255 by #1\char\count255}
\def\Alphabetic#1{\count255='100 \advance\count255 by #1\char\count255}
\def\Roman#1{\uppercase\expandafter{\romannumeral #1}}
\def\roman#1{\romannumeral #1}
\def\Number#1{\number #1}
\def\unnumberedchapters{\let\makel@bel=\relax \let\chapterlabel=\relax
\let\sectionlabel=\relax \equanumber=-1 }
\def\titlestyle#1{\par\begingroup \interlinepenalty=9999
     \leftskip=0.02\hsize plus 0.23\hsize minus 0.02\hsize
     \rightskip=\leftskip \parfillskip=0pt
     \hyphenpenalty=9000 \exhyphenpenalty=9000
     \tolerance=9999 \pretolerance=9000
     \spaceskip=0.333em \xspaceskip=0.5em
     \iftwelv@\fourteenpoint\else\twelvepoint\fi
   \noindent #1\par\endgroup }
\def\spacecheck#1{\dimen@=\pagegoal\advance\dimen@ by -\pagetotal
   \ifdim\dimen@<#1 \ifdim\dimen@>0pt \vfil\break \fi\fi}
\def\chapter#1{\par \penalty-300 \vskip\chapterskip
   \spacecheck\chapterminspace
   \chapterreset \titlestyle{\chapterlabel \ #1}
   \nobreak\vskip\headskip \penalty 30000
   \wlog{\string\chapter\ \chapterlabel} }

\def\section#1{\par \ifnum\the\lastpenalty=30000\else
   \penalty-200\vskip\sectionskip \spacecheck\sectionminspace\fi
   \wlog{\string\section\ \chapterlabel \the\sectionnumber}
   \global\advance\sectionnumber by 1  \noindent
   {\caps\enspace\chapterlabel \sectionlabel #1}\par
   \nobreak\vskip\headskip \penalty 30000 }
\def\subsection#1{\par
   \ifnum\the\lastpenalty=30000\else \penalty-100\smallskip \fi
   \noindent\undertext{#1}\enspace \vadjust{\penalty5000}}

\def\undertext#1{\vtop{\hbox{#1}\kern 1pt \hrule}}
\def\APPENDIX#1#2{\par\penalty-300\vskip\chapterskip
   \spacecheck\chapterminspace \chapterreset \xdef\chapterlabel{#1}
   \titlestyle{APPENDIX #2} \nobreak\vskip\headskip \penalty 30000
   \wlog{\string\Appendix\ \chapterlabel} }
\def\Appendix#1{\APPENDIX{#1}{#1}}
\def\appendix{\APPENDIX{A}{}}
%
%
%
\def\eqname#1{\relax \ifnum\the\equanumber<0
     \xdef#1{{\rm(\number-\equanumber)}}\global\advance\equanumber by -1
    \else \global\advance\equanumber by 1
      \xdef#1{{\rm(\chapterlabel \number\equanumber)}} \fi}
\def\eqinsert#1{\noalign{\dimen@=\prevdepth \nointerlineskip
   \setbox0=\hbox to\displaywidth{\hfil #1}
   \vbox to 0pt{\vss\hbox{$\!\box0\!$}\kern-0.5\baselineskip}
   \prevdepth=\dimen@}}
%

%

%

%
%
\def\GENITEM#1;#2{\par \hangafter=0 \hangindent=#1
    \Textindent{$ #2 $}\ignorespaces}
\outer\def\newitem#1=#2;{\gdef#1{\GENITEM #2;}}
\newdimen\itemsize                \itemsize=30pt
\newitem\item=1\itemsize;
\newitem\sitem=1.75\itemsize;     
\newitem\ssitem=2.5\itemsize;     
\outer\def\newlist#1=#2&#3&#4;{\toks0={#2}\toks1={#3}%
   \count255=\escapechar \escapechar=-1
   \alloc@0\list\countdef\insc@unt\listcount     \listcount=0
   \edef#1{\par
      \countdef\listcount=\the\allocationnumber
      \advance\listcount by 1
      \hangafter=0 \hangindent=#4
      \Textindent{\the\toks0{\listcount}\the\toks1}}
   \expandafter\expandafter\expandafter
    \edef\c@t#1{begin}{\par
      \countdef\listcount=\the\allocationnumber \listcount=1
      \hangafter=0 \hangindent=#4
      \Textindent{\the\toks0{\listcount}\the\toks1}}
   \expandafter\expandafter\expandafter
    \edef\c@t#1{con}{\par \hangafter=0 \hangindent=#4 \noindent}
   \escapechar=\count255}
\def\c@t#1#2{\csname\string#1#2\endcsname}
\newlist\point=\Number&.&1.0\itemsize;
\newlist\subpoint=(\alphabetic&)&1.75\itemsize;
\newlist\subsubpoint=(\roman&)&2.5\itemsize;
\newlist\cpoint=\Roman&.&1.0\itemsize;
%

%
%
%
\newcount\referencecount     \referencecount=0
\newif\ifreferenceopen       \newwrite\referencewrite
\newtoks\rw@toks
\def\NPrefmark#1{\attach{\scriptscriptstyle [ #1 ] }}
\let\PRrefmark=\attach
\def\CErefmark#1{\attach{\scriptstyle  #1 ) }}
\def\refmark#1{\relax\ifPhysRev\PRrefmark{#1}\else\NPrefmark{#1}\fi}
\def\crefmark#1{\relax\CErefmark{#1}}
\def\refend{\refmark{\number\referencecount}}
\newcount\lastrefsbegincount \lastrefsbegincount=0
\def\refsend{\refmark{\count255=\referencecount
   \advance\count255 by-\lastrefsbegincount
   \ifcase\count255 \number\referencecount
   \or \number\lastrefsbegincount,\number\referencecount
   \else \number\lastrefsbegincount-\number\referencecount \fi}}
\def\crefsend{\crefmark{\count255=\referencecount
   \advance\count255 by-\lastrefsbegincount
   \ifcase\count255 \number\referencecount
   \or \number\lastrefsbegincount,\number\referencecount
   \else \number\lastrefsbegincount-\number\referencecount \fi}}
\def\refch@ck{\chardef\rw@write=\referencewrite
   \ifreferenceopen \else \referenceopentrue
   \immediate\openout\referencewrite=referenc.texauxil \fi}
%
{\catcode`\^^M=\active 
  \gdef\obeyendofline{\catcode`\^^M\active \let^^M\ }}%
%
{\catcode`\^^M=\active 
  \gdef\ignoreendofline{\catcode`\^^M=5}}
{\obeyendofline\gdef\rw@start#1{\def\t@st{#1} \ifx\t@st\blankend%
\endgroup \@sf \relax \else \ifx\t@st\bl@nkend \endgroup \@sf \relax%
\else \rw@begin#1
\backtotext
\fi \fi } }
{\obeyendofline\gdef\rw@begin#1
{\def\n@xt{#1}\rw@toks={#1}\relax%
\rw@next}}
\def\blankend{}
{\obeylines\gdef\bl@nkend{
}}
\newif\iffirstrefline  \firstreflinetrue
\def\rwr@teswitch{\ifx\n@xt\blankend \let\n@xt=\rw@begin %
 \else\iffirstrefline \global\firstreflinefalse%
\immediate\write\rw@write{\noexpand\obeyendofline \the\rw@toks}%
\let\n@xt=\rw@begin%
      \else\ifx\n@xt\rw@@d \def\n@xt{\immediate\write\rw@write{%
        \noexpand\ignoreendofline}\endgroup \@sf}%
             \else \immediate\write\rw@write{\the\rw@toks}%
             \let\n@xt=\rw@begin\fi\fi \fi}
\def\rw@next{\rwr@teswitch\n@xt}
\def\rw@@d{\backtotext} \let\rw@end=\relax
\let\backtotext=\relax

\newdimen\refindent     \refindent=30pt
\def\refitem#1{\par \hangafter=0 \hangindent=\refindent \Textindent{#1}}
\def\REFNUM#1{\space@ver{}\refch@ck \firstreflinetrue%
 \global\advance\referencecount by 1 \xdef#1{\the\referencecount}}
\def\refnum#1{\space@ver{}\refch@ck \firstreflinetrue%
 \global\advance\referencecount by 1 \xdef#1{\the\referencecount}\refend}

\def\REF#1{\REFNUM#1%
 \immediate\write\referencewrite{%
 \noexpand\refitem{#1.}}%
\begingroup\obeyendofline\rw@start}
\def\ref{\refnum\?%
 \immediate\write\referencewrite{\noexpand\refitem{\?.}}%
\begingroup\obeyendofline\rw@start}
\def\Ref#1{\refnum#1%
 \immediate\write\referencewrite{\noexpand\refitem{#1.}}%
\begingroup\obeyendofline\rw@start}
\def\REFS#1{\REFNUM#1\global\lastrefsbegincount=\referencecount
\immediate\write\referencewrite{\noexpand\refitem{#1.}}%
\begingroup\obeyendofline\rw@start}
\newcount\figurecount     \figurecount=0
\newif\iffigureopen       \newwrite\figurewrite
\def\figch@ck{\chardef\rw@write=\figurewrite \iffigureopen\else
   \immediate\openout\figurewrite=figures.texauxil
   \figureopentrue\fi}
\def\FIGNUM#1{\space@ver{}\figch@ck \firstreflinetrue%
 \global\advance\figurecount by 1 \xdef#1{\the\figurecount}}
\def\FIG#1{\FIGNUM#1
   \immediate\write\figurewrite{\noexpand\refitem{#1.}}%
   \begingroup\obeyendofline\rw@start}
\def\fig{\FIGNUM\? fig.~\?%
\immediate\write\figurewrite{\noexpand\refitem{\?.}}%
\begingroup\obeyendofline\rw@start}
\def\figure{\FIGNUM\? figure~\?
   \immediate\write\figurewrite{\noexpand\refitem{\?.}}%
   \begingroup\obeyendofline\rw@start}
\def\Fig{\FIGNUM\? Fig.~\?%
\immediate\write\figurewrite{\noexpand\refitem{\?.}}%
\begingroup\obeyendofline\rw@start}
\def\Figure{\FIGNUM\? Figure~\?%
\immediate\write\figurewrite{\noexpand\refitem{\?.}}%
\begingroup\obeyendofline\rw@start}
\newcount\tablecount     \tablecount=0
\newif\iftableopen       \newwrite\tablewrite
\def\tabch@ck{\chardef\rw@write=\tablewrite \iftableopen\else
   \immediate\openout\tablewrite=tables.texauxil
   \tableopentrue\fi}
\def\TABNUM#1{\space@ver{}\tabch@ck \firstreflinetrue%
 \global\advance\tablecount by 1 \xdef#1{\the\tablecount}}
\def\TABLE#1{\TABNUM#1
   \immediate\write\tablewrite{\noexpand\refitem{#1.}}%
   \begingroup\obeyendofline\rw@start}
\def\Table{\TABNUM\? Table~\?%
\immediate\write\tablewrite{\noexpand\refitem{\?.}}%
\begingroup\obeyendofline\rw@start}
%

%
%
%
\def\masterreset{\global\pagenumber=1 \global\chapternumber=0
   \ifnum\the\equanumber<0\else \global\equanumber=0\fi
   \global\sectionnumber=0
   \global\referencecount=0 \global\figurecount=0 \global\tablecount=0 }
\def\FRONTPAGE{\ifvoid255\else\vfill\penalty-2000\fi
      \masterreset\global\frontpagetrue
      \global\lastf@@t=0 \global\footsymbolcount=0}

\def\paperstyle{\letterstylefalse\normalspace\papersize}
\def\letterstyle{\letterstyletrue\singlespace\lettersize}
\def\papersize{\hsize=35pc\vsize=48pc\hoffset=1pc\voffset=6pc
               \skip\footins=\bigskipamount}
\def\lettersize{\hsize=6.5in\vsize=8.5in\hoffset=0in\voffset=1in
   \skip\footins=\smallskipamount \multiply\skip\footins by 3 }
\paperstyle   
%
%
\def\MEMO{\letterstyle\FRONTPAGE \letterfrontheadline={\hfil}
    \line{\quad\fourteenrm FNAL MEMORANDUM\hfil\twelverm\the\date\quad}
    \medskip \memod@f}

\def\memit@m#1{\smallskip \hangafter=0 \hangindent=1in
      \Textindent{\caps #1}}
\def\memod@f{\xdef\to{\memit@m{To:}}\xdef\from{\memit@m{From:}}%
     \xdef\topic{\memit@m{Topic:}}\xdef\subject{\memit@m{Subject:}}%
     \xdef\rule{\bigskip\hrule height 1pt\bigskip}}
\memod@f
\newskip\lettertopfil
\lettertopfil = 0pt plus 1.5in minus 0pt
\newskip\letterbottomfil
\letterbottomfil = 0pt plus 2.3in minus 0pt
\newskip\spskip \setbox0\hbox{\ } \spskip=-1\wd0
\def\addressee#1{\medskip\rightline{\the\date\hskip 30pt} \bigskip
   \vskip\lettertopfil
   \ialign to\hsize{\strut ##\hfil\tabskip 0pt plus \hsize \cr #1\crcr}
   \medskip\noindent\hskip\spskip}
\newskip\signatureskip       \signatureskip=40pt
\def\signed#1{\par \penalty 9000 \bigskip \dt@pfalse
  \everycr={\noalign{\ifdt@p\vskip\signatureskip\global\dt@pfalse\fi}}
  \setbox0=\vbox{\singlespace \halign{\tabskip 0pt \strut ##\hfil\cr
   \noalign{\global\dt@ptrue}#1\crcr}}
  \line{\hskip 0.5\hsize minus 0.5\hsize \box0\hfil} \medskip }

\def\endletter{\ifnum\pagenumber=1 \vskip\letterbottomfil\supereject
\else \vfil\supereject \fi}
\newbox\letterb@x
\def\lettertext{\par\unvcopy\letterb@x\par}
\def\multiletter{\setbox\letterb@x=\vbox\bgroup
      \everypar{\vrule height 1\baselineskip depth 0pt width 0pt }
      \singlespace \topskip=\baselineskip }
\def\letterend{\par\egroup}
%
%
%
\newskip\frontpageskip
\newtoks\pubtype
\newtoks\Pubnum
\newtoks\pubnum
\newif\ifp@bblock  \p@bblocktrue
\def\PH@SR@V{\doubl@true \baselineskip=24.1pt plus 0.2pt minus 0.1pt
             \parskip= 3pt plus 2pt minus 1pt }
\def\PHYSREV{\paperstyle\PhysRevtrue\PH@SR@V}
\def\titlepage{\FRONTPAGE\paperstyle\ifPhysRev\PH@SR@V\fi
   \ifp@bblock\p@bblock\fi}
\def\nopubblock{\p@bblockfalse}

\frontpageskip=1\medskipamount plus .5fil
\pubtype={\tensl Preliminary Version}
\pubnum={0000}
\def\p@bblock{\begingroup \tabskip=\hsize minus \hsize
   \baselineskip=1.5\ht\strutbox \topspace-2\baselineskip
   \halign to\hsize{\strut ##\hfil\tabskip=0pt\crcr
   \the\Pubnum\cr \the\date\cr}\endgroup}

%
\def\title#1{\vskip\frontpageskip \titlestyle{#1} \vskip\headskip }
\def\author#1{\vskip\frontpageskip\titlestyle{\twelvecp #1}\nobreak}

\def\address#1{\par\kern 5pt\titlestyle{\twelvepoint\it #1}}
\def\andaddress{\par\kern 5pt \centerline{\sl and} \address}

\def\abstract{\vskip\frontpageskip\centerline{\fourteenrm ABSTRACT}
              \vskip\headskip }

%
%
%

\def\\{\relax\ifmmode\backslash\else$\backslash$\fi}
\def\globaleqnumbers{\relax\ifnum\the\equanumber<0%
\else\global\equanumber=-1\fi}

\def\journal#1&#2(#3){\unskip, \sl #1~\bf #2 \rm (19#3) }

\def\topspace{\hrule height 0pt depth 0pt \vskip}

\let\int=\intop         
\def\prop{\mathrel{{\mathchoice{\pr@p\scriptstyle}{\pr@p\scriptstyle}{
                \pr@p\scriptscriptstyle}{\pr@p\scriptscriptstyle} }}}
\def\pr@p#1{\setbox0=\hbox{$\cal #1 \char'103$}
   \hbox{$\cal #1 \char'117$\kern-.4\wd0\box0}}
\def\lsim{\mathrel{\mathpalette\@versim<}}
\def\gsim{\mathrel{\mathpalette\@versim>}}
\def\@versim#1#2{\lower0.2ex\vbox{\baselineskip\z@skip\lineskip\z@skip
  \lineskiplimit\z@\ialign{$\m@th#1\hfil##\hfil$\crcr#2\crcr\sim\crcr}}}
\def\leftrightarrowfill{$\m@th \mathord- \mkern-6mu
	\cleaders\hbox{$\mkern-2mu \mathord- \mkern-2mu$}\hfil
	\mkern-6mu \mathord\leftrightarrow$}
\def\lrover#1{\vbox{\ialign{##\crcr
	\leftrightarrowfill\crcr\noalign{\kern-1pt\nointerlineskip}
	$\hfil\displaystyle{#1}\hfil$\crcr}}}
%
%
%
\let\sec@nt=\sec
\def\sec{\relax\ifmmode\let\n@xt=\sec@nt\else\let\n@xt\section\fi\n@xt}
\def\obsolete#1{\message{Macro \string #1 is obsolete.}}
\def\firstsec#1{\obsolete\firstsec \section{#1}}
\def\firstsubsec#1{\obsolete\firstsubsec \subsection{#1}}
\def\thispage#1{\obsolete\thispage \global\pagenumber=#1\frontpagefalse}
\def\thischapter#1{\obsolete\thischapter \global\chapternumber=#1}
\def\nextequation#1{\obsolete\nextequation \global\equanumber=#1
   \ifnum\the\equanumber>0 \global\advance\equanumber by 1 \fi}
\def\BOXITEM{\afterassigment\B@XITEM\setbox0=}
\def\B@XITEM{\par\hangindent\wd0 \noindent\box0 }
%

%
\catcode`@=12 
\message{ by V.K.}
\everyjob{\input myphyx }
\def\etal{{\it et al.}}
\hsize=5.5truein
\voffset=-.3truein
\nopagenumbers
\baselineskip=14pt
\overfullrule=0pt
\def\ann{<\sigma v>_{ann}}
\def\msun{M_\odot}
\def\neut{\chi}
\REF\zwicky{Zwicky, F. 1933. Helvetica Physica Acta, \bf 11, \rm 110.}
\REF\freeman{Freeman, K.C. 1970. \sl ApJ. \bf 160, \rm 811.}
\REF\ostriker{Ostriker, J.P. \etal, 1974. \sl ApJ. Lett. \bf 193, \rm L1.}
\REF\faber{Faber, S.M. \& Gallagher, J.S. 1979. \sl Ann. Rev. Astron. Astroph.
	\bf 17, \rm 135.}
\REF\sancisi{Sancisi, R. \& Van Albada, T.S. 1986, in \sl
	Dark Matter in the Universe, IAU Symp. No. 117, \rm Knapp, G. \&
	Kormendy, eds., p. 67 (Reidel, Dordrecht).}
\REF\trimble{Trimble, V. 1987. \sl Ann. Rev. Aston. Astroph. \bf 25, \rm 425.}
\REF\sadoulet{Primack, J.R., Sadoulet, B., \& Seckel, D. 1988. \sl
	Ann. Rev. Nucl. Part. Phys. \bf B38, \rm 751.}
\REF\walker{Walker, T.P., \etal, 1991. \sl ApJ. \bf 376, \rm 51.}
\REF\mond{Begeman, K.G., Broeils, A.H., \& Sanders, R.H. 1991. \sl
	M.N.R.A.S. \bf 249, \rm 532.}
\REF\Tremaine{Tremaine, S. \& Gunn, J. 1979. \sl Phys. Rev. Lett. \bf 42,
	\rm 407.}
\REF\spergel{Gerhard, O.E. \& Spergel, D.N. 1992, \sl ApJ. \bf 389, \rm L9.}
\REF\burrows{Burrows, A., Klein, D., \& Gandhi, R. 1992,
	\sl Phys. Rev. \bf D45, \rm 3361.}
\REF\fitch{Fich, M. \& Tremaine, S. 1991, \sl Ann. Rev. Astron. Astroph.
	\bf 29, \rm 409.}
\REF\vanbiber{Van Bibber, K., private communication.}
\REF\sikivie{Sikivie, P. 1983. \sl Phys. Rev. Lett. \bf 51, \rm 1415.}
\REF\axion{for example, Turner, M.S. 1990, \sl Physics Reports \bf 197, \rm 67;
	Raffelt, G.G. 1990, \sl Physics Reports \bf 198, \rm 1.}
\REF\paczynski{Paczynski, B. 1986. \sl ApJ. \bf 304, \rm 1.}
\REF\griest{Griest, K. 1991. \sl ApJ. \bf 366, \rm 412.}
\REF\gould{Gould, A. 1992, \sl ApJ. \bf 392, \rm 234.}
\REF\alcocketal{Alcock, C. \etal, 1991, in \sl Robotic Telscopes of the
	1990's, \rm ed. Filippenko, A.V. (ASP, San Fransisco).}
\REF\french{Moscoso, L., \etal, 1991, preprint Saclay-DPhPE91-81;
	Spiro, M., private communication.}
\REF\ogle{Udalski, A. \etal, 1992. \sl Acta Astronomica \bf 42, \rm 253;
	Paczynski, B., private communication.}
\REF\kolb{Kolb, E.W. \& Turner, M.S. 1990. \sl The Early Universe, \rm
	(Addison-Wesley, Redwood City, California).}
\REF\griestsadoulet{Griest, K. \& Sadoulet, B., in \sl Dark Matter in the
	Universe, \rm eds. Galeotti, P., \& Schramm, D.N. (Kluwer, Netherlands,
	1989).}
\REF\griestsilk{Griest, K. \& Silk, J. 1990. \sl Nature \bf 343, \rm 26.}
\REF\goodman{Goodman, W.E. \& Witten. E. 1985. \sl Phys. Rev. \bf D31, \rm
	3059.}
\REF\silk{Silk, J., Olive, K.A., \& Srednicki, M. 1985. \sl Phys. Rev. Lett.
	\bf 55, \rm 259.}
\REF\indirect{For example,
	Mori, M. \etal, 1991. \sl Phys. Lett. \bf B270, \rm 89;
	Sato, N. \etal, 1991. \sl Phys. Rev. \bf D44, \rm 2220;
	Barwick, S. \etal, 1992. \sl J. Phys. \bf G18, \rm 225.}
\REF\betty{Young, B.A, 1992, in Proceedings of the Third Intl. Conf. on
	Advanced Technology and Particle Physics, Borchi, E., \etal, eds.
	(Elsevier).}
\REF\ellishag{Ellis, J., Hagelin, J.S., Nanopoulos, D.V., Olive, K.A.,
	Srednicki, M. 1984. \sl Nucl. Phys. \bf B238, \rm 453.}
\REF\haber{Haber, H.E. \& Kane, G.L. 1985. \sl Physics Reports \bf 117, \rm
75.}
\REF\griestros{Griest, K. \& Roszkowski, L. 1992. \sl Phys. Rev.
	\bf D46, \rm 3309.}
\REF\shutt{Shutt, T. \sl et al. \rm 1992.  \sl Phys. Rev. Lett. \bf 69,
	\rm 3425; \sl ibid \rm 3531.}
\REF\kamion{Halzen, F., Stelzer, T., \& Kamionkowski, M. 1992, \sl Phys.
	Rev. \bf D45, \rm 4439.}
\REF\griestneut{Griest, K. 1988. \sl Phys. Rev. \bf D38, \rm 2357.}
\REF\minimalsugra{for example,
	Nojiri, M.M. 1991, \sl Phys. Lett. \bf B261, \rm 76;
	Ellis, J. \& Roszkowski, L. 1992, \sl Phys. Lett. \bf B283, \rm 252;
	Ross, G.G. \& Roberts, R.G. 1992, \sl Nucl. Phys. \bf B377, \rm 571;
	Drees, M \& Nojiri, M.M. 1992, \sl Phys. Rev. \bf D45, \rm 2482;
	Lopez, J.L., Nanopoulos, D.V., \& Yuan, K. 1992,
		\sl Nucl. Phys. \bf B370, \rm 445;
	Drees, M. \& Nojiri, M.M. 1992, preprint DESY-92-101;
	Roszkowski, L. 1992, private communication.}
\REF\dreestata{Drees, M. \& Tata, X. 1991, \sl Phys. Rev. \bf D43, \rm 2971.}
\REF\ibanez{Ibanez, L.E., \& Lust, D. 1992, \sl Nucl. Phys. \bf B382, \rm 305.}
\REF\randall{Hall, L.J., \& Randall, L. 1990, \sl Phys. Rev. Lett.
	\bf 65, \rm 2939.}
\REF\extendedsusy{For example, Flores, R.A., Olive, K.A., and Thomas, D. 1991,
	\sl Phys. Lett. \bf B245, \rm 514;
	Abel, S.A., Cottingham, W.N., \& Whittingham, I. 1990, \sl
	Phys. Lett. \bf B244, \rm 327;
	Flores, R.A., Olive, K.A., \& Thomas, D. 1991, \sl Phys. Lett.
	\bf B263, \rm 425.}
\font\Bigbf=cmbx12 scaled 1440
\voffset= -.3truein
\nopagenumbers
\line{\hfil UCSD/ASTRO-TH 93-01}
\line{\hfil Janurary 1993}
\vskip .75truein
\centerline{\Bigbf The Search for Dark Matter:}
\smallskip
\centerline{\Bigbf WIMPs and MACHOs
\foot{
Invited review talk presented at the Texas/PASCOS Symposium,
Berkeley, CA, Dec. 1992; to appear in the Proceedings.}}
\vskip 1.0truein
\centerline{\bf Kim Griest}
\vskip .5truein
\centerline{University of California, San Diego}
\centerline{Physics Department-0319}
\centerline{La Jolla, CA 92093}
\bigskip\bigskip
\centerline{\bf Abstract}
\smallskip
We review the status of experiments and ideas relevant for the detection
of the dark matter which is suspected to be the dominant constituent of
the Universe.  Great progress is being made and the chances are non-negligible
that one of the many currently in-progress experiments will discover the
nature of the dark matter.  We discuss the main dark matter candidates,
and review the experiments relevant to each of them.
\vfill
\eject
\pagenumbers
\centerline{\bf INTRODUCTION}
This is an extremely interesting time for dark matter (DM) detection.
I think the chances are fair that within the next few years, we
will finally know what the primary constituent of the Universe is.
Of course there is no guarantee of this;  the dark matter could
consist of something which interacts only gravitationally and will never
be discovered.
But in this talk I would like to describe the
ideas and experiments which lead me to believe that this may not
be the case; the primary reason being that for the first time experiments
are being built which will have the sensitivity to
detect some of the favorite dark matter candidates.

You might be surprised at my optimism given that many talks on dark matter
begin by saying that the dark matter problem has been around since 1933,
when Zwicky
\refmark\zwicky
measured the velocities of galaxies in several clusters
and showed that there was much more mass in the clusters than could be
accounted for by the stars in the galaxies themselves.  After all,
if so little progress has been made in the last 60 years, why are the
next few so special?
This attitude, however, is a little misleading.
First, Zwicky's work did not get a lot of attention at the time, and
it really wasn't until the 1970's, that the idea that dark
matter dominates the halos of galaxies like our own became accepted.
\refmark{\freeman,\ostriker,\faber}
In fact, it is really only during
the last 10-15 years, that the really incontrovertible data
(such as neutral hydrogen rotation curves of spiral galaxies)
has become available, and that a consensus has been reached
that the dark matter
problem is one of the major unsolved problems in physics.
\refmark{\sancisi,\trimble}
Second,
major efforts at detecting the dark matter are even more recent.
\refmark\sadoulet
It is remarkable, that while various
pilot experiments, etc. have been going for a while,
it is only within the last 2-3 years that experiments capable of detecting
realistic dark matter candidates have been conceived and funded.
And these experiments have yet to turn in results.  However, before
discussing these experiments, let me remind you which of the several
dark matter problems I am talking about.

The mass density of the Universe $\Omega$ is measured in units of the critical
density $\rho_{crit}$, and the luminous matter (stars, dust, and gas)
contribute
$\Omega$ between roughly 0.005 and 0.007.
The most secure dark matter
is that in the halos of spiral galaxies, which contribute $\Omega$
between roughly 0.03 and 0.1.  It is the gap between these numbers
which constitutes the most robust dark matter problem, and it is
precisely this halo dark matter for which the detection experiments search.
On larger scales, groups and clusters of galaxies contribute
perhaps $\Omega$ between 0.05 and .3, but as we have seen in the
talk of Mushotsky (these proceeding),
the cosmic virial theorem used to make these
estimates is not completely foolproof.  On even larger scales,
we heard Bertshinger (these proceedings)
describe the impressive agreement
between the potential field derived from peculiar velocities
using the POTENT method, and the source counts of IRAS selected galaxies.
This agreement implies $\Omega^{.6}/b = 1.2 \pm .6$, giving a value
of $\Omega$ near unity for any reasonable value of the bias factor $b$.
However, this technique is still somewhat controversial, and I think it
is too soon to declare $\Omega=1$ the answer, even though inflationary
cosmology and aesthetics prefer this value.
Finally, big bang nucleosynthesis is consistent with the measured abundances
of light elements only if baryons exist with a density of
$0.01< \Omega_{baryon} < 0.1$,
\refmark\walker
where I have taken somewhat larger
error bars than Tonry (these proceedings)
mostly because I am less sure
of the value of the Hubble constant $h$.
There are several points
to note from this mass density inventory.  First, it is entirely possible,
that ALL the really secure dark matter, the dark matter in spiral galaxies,
consists of baryons.  To dismiss the searches for baryonic
dark matter and concentrate solely on the search for
particle dark matter is premature at best, and may result in missing
the dark matter all together.  Next, IF $\Omega=1$, it is clear that
all the dark matter cannot be baryonic.  In this case, the bulk of
the mass density of the Universe must consist of some elementary particle
(or non-Newtonian gravity, etc. which I will not discuss here!).
Finally, the dark matter problem exists entirely independent of whether
or not $\Omega=1$, and whether or not one believes the nucleosynthesis
limits.  In fact, the dark matter searches are looking ONLY
for the dark matter in the Milky Way, which, luckily, is quite secure.
\bigskip
\centerline{\bf THE CANDIDATES}
But how does one search for something which neither emits nor absorbs
electromagnetic radiation in any known waveband?  It can be discouraging,
especially since serious candidates have been suggested which span some
70 orders of magnitude in mass, from $10^{-5}$eV axions to
$10^6 M_\odot$ black holes.  It is clear that no
one technique can work for such a diverse set of candidates.
One must pick a candidate
and design a search technique for it.  Luckily, however, some candidates
are ``generic" and ``generic" techniques can cover a class of candidates.
Recently, many classes of candidates are becoming experimentally
accessible for the first time.

\topinsert
$$\vbox{ \halign{ #\hfil && \quad # \hfil \cr
\noalign{\hrule} & \cr
{\bf Candidate } & {\bf Examples} \cr
\noalign{\hrule} & \cr
axion & DFS, Hadronic \cr
WIMP & neutralino, GeV neutrino, technicolor \cr
& \qquad particle, extra Higgs,  etc., etc., etc. \cr
MACHO & jupiters, brown dwarfs,\cr
&\qquad  black hole remnants \cr
light neutrino & $\nu_\mu$ or $\nu_\tau$ with 30 eV mass\cr
non-Newtonian gravity & Milgrom's MOND\refmark\mond\cr
none-of-the-above & ? \cr
\noalign{\hrule} & \cr
}}$$
\centerline{\bf Table 1. \rm Dark Matter Candidates}
\endinsert
In Table 1, I've listed some of the main classes of candidates.
The axion is in a class by itself, and needs special detection techniques.
The largest class of candidates is the Weakly Interacting Massive
Particle (WIMP)
group, which consists of literally hundreds of suggested particles,
the supersymmetric neutralino particle being the current favorite.
Another large group is the Massive Astrophysical Compact Halo Object (MACHO)
group, which consists of balls of hydrogen and helium too light to
initiate nuclear burning (jupiters at 0.001 $\msun$, brown dwarfs at 0.01
$\msun$, etc.), as well as the massive black hole
remnants of an early generation of
stars.  These can be searched for by gravitational microlensing.
I won't discuss non-Newtonian gravity, or the dark horse favorite
``none-of-the-above" candidate.  Also, since it is neither a WIMP nor
a MACHO, I won't say much about light
neutrinos.  It is possible for the halo of our galaxy to
consist of 30 eV tau neutrinos.  However, such a neutrino cannot
make-up ALL of the dark matter, since phase-space constraints
\refmark\Tremaine
show that such a light fermion cannot make-up the halos of the small dwarf
galaxies, which are known to contain large amounts of dark matter.
\refmark\spergel
So at least one other type of dark matter would be needed.  Also, if our
halo did consist of light neutrinos, there is currently no good idea of how to
detect them;  they carry too little momentum and energy.
The only possibility would be to measure the neutrino mass and use
the big bang calculation
$ \Omega_\nu \approx {m_\nu/(100 h^2)}$,
where $0.5 \leq h \leq 1$ parameterizes our ignorance of the Hubble parameter.
While a direct measurement of the tau neutrino mass is probably not
possible, the situation
is not completely hopeless, since
IF $\nu_\tau-\nu_\mu$ mixing is large, the CERN and/or
Fermilab neutrino oscillation experiments might be able to
determine the mass and mixing.
The only other alternative I know of is if there is a supernova
in our galaxy, then the Sudbury Neutrino Observatory might be able to
separate the tau neutrinos from the mu and e neutrinos,
measure the time delay, and therefore
measure $m_\nu$.
\refmark\burrows
But this looks like a bit of a long shot.
Finally, the recently popular mixed hot plus cold dark matter scenario
of galaxy formation,
which seems to survive all the large and small scale structure tests
as well the microwave background measurements, hypothesizes
$\Omega=1$ with
70\% cold dark matter and 30\% hot dark matter.  The hot dark matter
would be in the form of a
7 eV tau neutrino and the cold component in the form
of some more massive new elementary particle.
While this scenario is certainly possible, it
is important to recall that such a light
neutrino would not cluster much on the galaxy scale, implying that
the dark matter in our halo would consist almost entirely of the
cold dark matter component.  Thus WIMP detection strategies would be
almost unchanged in the mixed DM scenario.

Next, let me remind you that while we don't know what the dark matter
is, we know fairly well where it is and how fast it is moving.  This is
important for the detection schemes.  The rotation curves determine
that the dark matter density
drops as $r^{-2}$ at large galactic radii $r$, and that the
velocity dispersion
is roughly 270 km/sec.  The halo is thought to be roughly spherical,
but there is not much evidence to support this.
The halo also probably extends out to at least 50 kpc from the center
of the galaxy.
\refmark\fitch
The dark matter density at the Earth's position in the galaxy
is roughly 0.03 GeV/cm$^3$ (0.007 $\msun/pc^3$).
Thus for particle dark matter matter detection schemes we know roughly
how many particles are passing through the laboratories, and how much
energy the particles carry (as a function of the particle mass).
For the microlensing schemes, we know roughly what the density and
velocity distributions are.  Thus reasonably accurate estimates of
detection rates are possible.

Now while I am not going to say much about axions, I can't resist
telling you about a new experiment which has just been funded.
\refmark\vanbiber
This is a collaboration of Lawrence Livermore Lab, the Institute for
Nuclear Research, the University of Florida, MIT, etc., to build a
very large superconducting magnet for the purpose of detecting dark
matter axions, if they make up the galactic halo.  The basic idea is
that since the axion couples to two photons, an axion can interact with
a strong magnetic field and convert into a photon which can be detected
as a resonance in a microwave cavity.
\refmark{\vanbiber,\sikivie,\axion}
If axions have masses in the
$10^{-6}$ to $10^{-5}$eV range (corresponding to frequencies in the
40-400 MHz range), they can contribute $\Omega \approx 1$ and be the
dark matter.
\refmark\axion
One slowly tunes the microwave cavity to be sensitive
to different axion masses.  Two previous experiments have run
and found no signal for axions, but they only had sensitivity to
a density of axions a factor of 100 greater than could possibly exist.
\refmark{\axion}
The exciting new development, is that due to a bigger magnet
and a lower noise system, this new LLNL experiment should, for the
first time, be sensitive to a halo density of realistic axions.
While it won't have the sensitivity to detect or rule-out all the
axion models (It can see much of the ``hadronic axion" parameter space,
but not the Dine-Fischler-Srednicki axion), it should be up an running
within two years, and could be improved to make a more definitive search.
\refmark\vanbiber

\bigskip
\centerline{\bf MACHO'S AND THEIR DETECTION}
I will now turn to the MACHO searches.  These are probably the most exciting
dark matter searches under way.  They are the only ones with a chance of being
definitive.  And whether these searches find DM or not, the results will
be important.  Remember, that ALL the DM in the Milky Way halo
could be baryonic, and still be consistent with big bang nucleosynthesis.
If it is baryonic, that is, consisting of
hydrogen and helium, it is most likely
in the form of jupiter-like planets, brown dwarf stars, or the black hole
remnants of an early generation of stars.  The microlensing searches
can give definitive results on dark objects in these mass ranges
($10^{-5}$ to $10^2 M_\odot$).
\refmark{\paczynski,\griest}
If the dark matter consists of these
objects, these experiments should find it;  and it it doesn't, these
experiments should prove it doesn't.  In fact, these searches may eventually
be sensitive to the entire theoretically possible range of baryonic DM
($10^{-9}$ to $10^6\msun$ (Ref. \gould)).
These experiments use an idea of Paczynski,
\refmark\paczynski
which is based on
an enormous amount of earlier work on gravitational lensing.
If a dark object lies directly on the line-of-sight between us (the observer)
and a distant star (the source), it forms a gravitational lens,
which bends the
starlight, making the star appear as a ring.  The radius of this Einstein
ring sets the scale for the experiment and is given by
$R_e = (2L^{-1/2}/c)[Gmx(L-x)]^{1/2},$
where $m$ is the mass of the lens, $L$ is the source-observer distance,
and $x$ is the lens-observer distance.  Note that $R_e$ is proportional
to the square-root of the lens mass.
Now if the halo consists of MACHOs, they occasionally will come close
to the line-of-sight to a star, and cause a microlensing event.
It is very rare, however, that a perfect alignment will occur.
If the alignment is not perfect, two stellar images appear rather than
a ring.  For distances appropriate for the MACHO searches, these images
are too close together to be resolved from the ground (milliarcseconds),
but the light from the two images add, making the star appear brighter.
The amplification (magnification) of the star is given by
$A = (u^2+2)/[u(u^2+4)^{1/2}]$, where $u = b/R_e$ is the distance
of the lens from the line-of-sight in units of $R_e$.
For a MACHO moving with transverse speed $v$, the ``lightcurve" is
given by $A(u(t))$, where $u(t) = [u^2_{min} + (t/{\hat t})^2]^{1/2}$,
and the time scale of the event is set by ${\hat t} = R_e/v$.
Thus, as a MACHO moves by, a star will appear to brighten, then return to
normal;  $A$ will increase from unity to a maximum value $A_{max}$, and then
decreases back to unity.  This brightening is symmetric in time and its
duration depends upon the MACHO speed and mass.
Since the halo density and velocity distributions are known roughly,
a calculation of how often such microlensing events take place, and
how long typical events last can be made \refmark{\paczynski, \griest}.
If one monitored 3 million stars nightly,
one expects expects around 170 events/yr for a halo density of jupiter mass
objects (0.001$\msun$), and an average event duration of about 3 days.
For brown dwarf mass dark matter (0.1$\msun$), one expects about 20 events/yr
with an average duration of about 30 days.  Lighter MACHOs give more events
with shorter durations.  In fact, the product of the event rate and
average duration is the optical depth for microlensing $\tau \approx
5 \times 10^{-7}$ and is independent of the MACHO mass.  (An event is defined
here as the time during which a star is 34\% brighter than normal.)
So by modifying the observing strategy one can get sensitivity
from $m\sim10^{-8}\msun$ to $m\sim10^2\msun$ or larger.
The lower limit comes because for small mass lenses the projected Einstein
ring radius becomes smaller than the stellar source radius, and large
amplifications become impossible.  The upper limit, which is very
rough, comes when the total duration of the event is longer than
the duration of the experiment.  (But see ref.~\gould, for possible
extensions of the mass range sensitivity.)

The obvious problem here is that it takes millions of stars to get
a handful of events.  And occurrence of variable stars of all types is
known to be much more frequent than this.  So there was (and still is!)
a lot of skepticism;  can such an experiment be done?
Well, three groups are doing it and already returning excellent data!
But what about backgrounds such as cosmic rays, satellite tracks,
variable stars, and new types of transient phenomena?  All these
must be identified and distinguished from the bona fide microlensing events
(which may or may not occur).

Luckily, bona fide microlensing has many powerful signatures, some of which
are listed below:
\pointbegin
Very high amplification events occur ($\Delta$ magnitude $> .75$), which
by normal astronomical standards are very easy to detect.
\point
The lightcurves have a unique shape.  Only 3 parameters $A_{max}$,
the duration $t_e$, and the time of peak, completely determine $A(t)$.
One can get many points on a lightcurve and check the shape.
\point
Microlensing is achromatic.  The lightcurve should be identical in both
(all) filter bands.  The experiments monitor the stars in at least two
colors to take advantage of this signature.  Note that most variable stars
change color as they vary.
\point
The distribution in peak amplification $A_{max}$ is known a priori.
The $A_{max}$ is determined solely by the distance of closest approach
to the line-of-sight, and the distribution of these distances is uniform.
\point
ALL signatures, distributions in $A_{max}$, $t_e$, etc. are independent
of the star type and luminosity.  In the LMC, giant stars, A type stars,
and B type stars will be monitored.  If a new type of variable star exists
which passes all the previous tests, but occurs only for stars of a certain
type, it cannot be microlensing.
\point
The microlensing probability is so small that it should occur only {\it once}
on a given star.  If a star which had a microlensing-like event continues
to vary, it was not microlensing.
\point
It is possible to catch microlensing in the act and notify other
observers.  For example, if an event with duration of
5 weeks were discovered at week 3, one could predict
the lightcurve for the next few weeks and get hundreds of points
to check the shape, etc.
\point
Finally, if the dark matter consists of MACHOs, microlensing rates should
be reproducible from year to year (with small seasonal modulations in rates
\refmark\griest
).
If 20 events are found one year, 20 similar events on random stars
should be found again the next year.

To demonstrate the power of these signatures, let me show you some
of the data from the experiment of which I am a member (the MACHO
collaboration
\refmark\alcocketal).
We take images of 10-20 million stars per night, and have performed preliminary
analysis on about 1 million stars.  About 800 of these passed
the first level microlensing trigger and were examined by eye.
(This trigger just finds stars with bumps in their lightcurves which are
well correlated in the red and blue filter bands.)

The lightcurves in Figure 1a,b,c look like reasonable microlensing
candidates at first glance, the shapes look not too far off microlensing;
but notice that they are very chromatic (that is,
have different peak variations in the red and blue bands, as is the case
with many variable stars).
Thus these are long period variable stars and not microlensing.  (Further
monitoring of these stars will be able to verify this.)
Of course, most first level candidates
don't have the right shape (e.g. Figures 1d, e).
Next,
consider Figure 1f, which is not unusual, but illustrates that our automated
telescope, photometry and analysis programs can take good data on, and find
stars such as this, which vary only by 20\%.  Events with amplifications of
100\% and more are expected to occur, and are easy to pick out with
data of this quality.
Finally, many variable stars
look like Figure 1g.  Is this a random scatter plot?
No, it is a variable star with a period which is
short compared to our sampling time.  When we find the
proper period and fold this data it becomes Figure 1h, a beautiful Cepheid
variable.

So let me summarize the status of three experiments I know about.
The MACHO collaboration consists of physicists and
astronomers from Livermore, the Center for Particle Astrophysics,
and Mt. Stromlo Observatory.
\refmark\alcocketal
They have exclusive use of a 1.3 m telescope on Mt. Stromlo for at least
4 years, and have taken about 180 nights of preliminary data.
They have a very large camera, consisting of 8 2K by 2K pixel CCD's,
for a total of 32M pixels.  Using a dichroic filter
they take exposures simultaneously in red and blue
with a field of about 0.5 deg$^2$.
They target both the LMC and the galactic bulge.  Each clear night
data is taken on 10-20$\times 10^6$ stars, with
2-3$\times 10^6$ having errors less than 0.1 mag.  They have done
a preliminary analysis on around 1 million stars, found roughly 2000
variable stars of all types, but have no clear microlensing event yet.

The French collaboration
\refmark\french
is also a large collection of physicists and astronomers from eight
or so French institutions.  They use a 0.4 meter telescope for
their CCD work and a Schmidt for the large 36 deg$^2$ photographic plates.
They took $\sim$ 2500 CCD images last year at ESO, and expect $\sim 7000$
this year.  They also have acquired around 300 Schmidt plates.
Their camera contains 16 400$\times$580 pixel CCD's for a total
of $\sim 3.7$ Mpixels and a 0.5 deg$^2$ field of view.  So far they have
targeted only the LMC.  They expect $3\times 10^6$ stars with errors less
than 0.15 mag from their plates, and get about 50,000 stars from
their CCD exposures.  They have found variable stars and a flare star,
but have no clear microlensing candidate yet.

The ``OGLE" collaboration
\refmark\ogle
consists of astronomers from several Polish institutions and the Carnegie
Institute,
and Bohdan Paczynski, the inventor of the idea.
They do not have their own telescope at this point, but took 45 nights
of data on a 1 meter telescope at Las Campanas
last year, and have been awarded 70
nights this year.  They use a 2K by 2K pixel CCD, (4Mpixels), with
a 0.06 deg$^2$ field of view.  They target the galactic bulge exclusively.
They have $\sim 0.75 \times 10^6$ stars, with errors $< .1$ mag.
They have found many variables, but as yet have no microlensing events.

So while none of the experiments have yet to present a detection of
or limits on baryonic on dark matter, I think it is remarkable
how quickly they have come on line, and I expect detections or
limits to appear soon.  Recall, that the idea appeared only in 1986, and the
first experiment was started only in 1990.

\bigskip
\centerline{\bf WIMPS AND THEIR DETECTION}
While I think the search for baryonic dark matter is important and very
exciting, it is good to remember, that no one has thought of a
good way to put 90\% of the mass of the Universe in such objects, and
that if $\Omega$ is near unity (as recent
observation suggests), the dark matter cannot consist entirely
of baryons.  Thus it is important to search for non-baryonic dark matter
in the form of new elementary particles.  This, at first glance, seems
hopeless.  The dark matter could consist of elementary particles
which have only gravitation interaction, and therefore are undetectable
in laboratory experiments.  However, there is a reason,
coming from the big bang, which suggests this may not be the case.

Consider any stable particle $\delta$ in equilibrium
in the Early Universe.  When the temperature is much higher
than its mass $m_\delta$, it
exists with a number density roughly equal to that of photons.
As the Universe cools and expands, its number density $n$ drops.
As the temperature finally
drops below the mass of the $\delta$, the $\delta$ becomes
much more difficult to create thermally, and its number density then
drops exponentially (Boltzmann factor: $n \sim e^{-m_\delta/T}$).
As the Universe continues to cool, this
exponential drop can not continue forever, however, because eventually
the number density becomes so low that there is almost no chance that
a $\delta$ and $\bar \delta$ meet and annihilate
(remember for $\delta$ to be the dark matter it must be stable).
Thus a certain number of $\delta$'s must survive the big bang
and still be around today.  How many survive?  It depends upon the
thermally averaged annihilation cross section $\ann$.
The calculation of the relic abundance
of particles from the Early Universe
has been done many times with varying degrees of
accuracy, but the result can be roughly summarized by
\refmark\kolb
$ \Omega_\delta h^2 \approx 10^{-27} {\rm cm}^3{\rm sec}^{-1} /\ann$.

Now for any particle to be a dark matter candidate it must have
a relic abundance $0.025 \leq \Omega_\delta h^2 \leq 1$,
where the upper bound comes from requiring
the age of the Universe to be at least
$10^{10}$ years, and the lower bound comes from
$\Omega = .1$ and $h=0.5$.
This equation says that ANY dark matter candidate which was once
in thermal equilibrium MUST have a self annihilation cross section
$\ann(\delta{\bar\delta}\rightarrow e^+e^-, q{\bar q},etc.)/c
\approx 10^{-36} {\rm cm}^2$.
This is the weak interaction strength.  Thus a particle which had only
gravitational interactions, would give far too high a relic abundance,
if it was ever in thermal equilibrium.  Therefore, one does not expect
particle dark matter to have either extremely feeble or extremely
strong interactions; it should interact with just the strength of
the weak interaction.
The fact that it is precisely the weak interaction strength which is
required of particle dark matter is a very interesting coincidence
(or is it a significant clue?).  Note that it is just in this range that
particle accelerators around the world are searching for new particles.
And particle theories which attempt to solve outstanding theoretical problems
with the Standard Model typically predict new particles
with just this interaction strength.  Keep in mind also that
since $\delta\delta\rightarrow e^+e^-, q{\bar q}$, etc. is of weak
strength, the inverse processes, $e^+e^-\rightarrow\delta{\bar\delta}$
used to search for particles in accelerators
will also be of this order of magnitude,
as will the ``crossed" elastic scattering process
$\delta q \rightarrow \delta q$, which the direct DM searches rely on.

Thus from the postulate that some WIMP from the early Universe
is the dark matter one can get rough predictions of its detectability,
both in particle accelerators, and in the underground germanium detectors,
etc.  Remarkably, the new generation of experiments are now probing precisely
the range of parameters where the dark matter should exist.
This is the primary reason for my guarded optimism regarding
dark matter searches.  However, it is important to keep in mind
that there are many exceptions to these generic predictions,
\refmark\griestsadoulet
and that it is possible to have DM particles which do not come from
equilibrium in the early Universe (e.g. the axion!), and therefore
the dark matter could consist of undetectable particles.
In specific models, where the relevant cross sections can be
calculated, these generic predictions can be checked and made more
precise, and in fact, usually are borne out.  For example, the first
particle cold dark matter candidate, the GeV mass Dirac neutrino
of Lee and Weinberg (among others), has now been ruled out by a combination
of the above ideas.
\refmark\griestsilk

So it should be clear that accelerators are great places
to look for dark matter, and if I
had to place a bet, I think CERN or the SSC are the places most likely
to find the dark matter.
However, there are at least 2 other promising experimental techniques
under rapid development which are
(or soon will be) exploring the same regions of parameter
space in a competitive manner.
The first is a direct detection technique,
\refmark\goodman
which attempts to
measure the small (O(keV)) energy deposited in a kilogram size crystal
when a dark matter particle elastically scatters off a nucleus.
These crystals are typically kept at cryogenic temperatures and in very
low radiation background environments (underground).  The second
class of experiments attempts indirect detection of dark matter particles
by measuring the neutrinos which result from $\delta + {\bar\delta}$
annihilation in the Sun or Earth.
\refmark\silk
In this scenario, dark matter particles collect in
the Sun or Earth over billions of years and gradually settle to the center.
Neutrinos from the dark matter particle
self-annihilation stream out of the Sun and could be
detected in proton decay experiments such as IMB, Kamiokande, and
soon in the newer detectors such as MACRO, AMANDA, and DUMAND.
\refmark\indirect
These neutrinos are high energy ($E_\nu \approx 0.5 m_\delta$)
and so are easier to detect than the normal solar neutrinos,
but there are fewer of them so large detectors are needed.

Now there have been hundreds of particles suggested as dark matter
candidates, but recently most attention has been focussed on
the lightest supersymmetric particle (LSP) and the most likely LSP is
the neutralino ($\neut$), which is also known as the photino, Higgsino,
B-ino, and Z-ino.
\refmark{\ellishag,\haber}
The neutralino is a linear combination of the
supersymmetric partners of the photon, Higgs boson, and Z boson.
The mass and couplings of the neutralino
depend upon 4 parameters: a Higgsino mass parameter $\mu$, two soft
SUSY breaking masses $M_1$ and $M_2$, and the ratio of Higgs boson
vacuum expectation values $\tan\beta$.  To calculate cross sections
one also needs the squark, slepton, and Higgs masses, as well as the
top quark mass.
Now while there have been dozens of WIMP candidates proposed, the
neutralino is the front runner for several reasons, and hundreds of
papers on neutralinos have been written, and many experimental searches
for neutralinos performed.  The motivations for supersymmetry (SUSY)
have been extensively discussed at this meeting (Zwirner, Ellis, Wilczek,
these proceedings), and basically have to do with incorporating
gravity into the Standard Model, solving the gauge hierarchy problem,
and allowing coupling constant unification to be consistent with new
LEP data.
While there are many models and many parameters in each model,
once these have been
specified, the cross sections and relic abundance can be calculated
and the generic predictions previously discussed made precise.

For example, Figure 2 shows a scatter plot of many parameter choices
in a specific minimal SUSY model.  I've plotted the relic abundance
vs. the neutralino mass.  Each cross represents a different choice
of SUSY parameters, and the density of crosses indicates the
amount of parameter space which predicts that value of the relic abundance.
To be DM, the relic abundance must be between 0.025 and 1, as indicated
by the horizontal lines.  It is apparent from the figure
that much of parameter space gives
relic abundances in the range relevant for dark matter.  There are
choices of parameters which give too little DM, and choices which give
too much and therefore are ruled out, but it is clear that
$\Omega \sim 1$ in neutralinos is a generic feature of the model,
and therefore adds to the attractiveness of the neutralino as a DM
candidate.
Also note in Figure 2,
that parameter choices which gave $m_\chi < \sim 20$ GeV are not shown,
since they are ruled out by the LEP experiments.
Thus prime regions of
parameter space for dark
matter candidates are being explored by the particle accelerators.

Let me now turn to the direct detection experiments.
There are dozens of groups around the world actively pursuing
these experiments, each one of which could give an interesting hour
long talk, so I can only list some of the experiments
\refmark\betty
and mention
one example briefly. (See Table 2.)
\topinsert
$$\vbox{ \halign{\hfil #\hfil && \quad\hfil #\hfil \cr
\noalign{\hrule} & \cr
{\bf Group} & {\bf Technique} & {\bf Material} \cr
\noalign{\hrule} & \cr
CfPA/UCB/UCSB,          & Ionization + phonons  & Ge, Si, GaAs \cr
Imperial Col.,          & via thermistor &              \cr
Saclay                  &                       &              \cr
Stanford,               & Transition edge of super-   & W on Si,     \cr
MPI Munich,             & conducting films & Sapphire  \cr
IAS/CNRS                &                       &              \cr
UBS,                     & superheated, super-          & Indium/Mylar, \cr
Annecy,                  & conducting granules   & Tin/Kapton    \cr
Interferometrics        & \quad                 & \cr
Milan                   & scintillation + phonons & CaF$_2$ \cr
Brown U.                & rotons                & He \cr
Saclay,                  & scintillation         & NaI \cr
Osaka/Beijing/Roma      &                       & \cr
\noalign{\hrule} & \cr
}}$$
\centerline{Table 2: Partial list of direct detection experiment
	groups\refmark\betty}
\endinsert

Now in all these experiments the problem is not really the rate,
but the background.  Taking the weak interaction strength, which comes
from the relic abundance calculation, and estimating the elastic scattering
cross section by crossing symmetry, one estimates an event
rate in the cryogenic detectors of about 1 event/kg/day.
\refmark\griestsadoulet
Thus a few
kilogram detector run for several months should have plenty of events.
Keep in mind, however, that there are uncertainties regarding the crossing,
and that in the detectors it is the WIMP-nucleus, not the WIMP-quark
interactions that are relevant.  This means one needs to add the quark
contributions to get the WIMP-nucleon contribution, and use nuclear
wave functions to add coherently the nucleon contributions.  Thus, the
actual rates may differ by several orders of magnitudes from the
generic rate.  For example, GeV mass Dirac neutrinos have a fortunate
coherence effect which result in an event rate of hundreds per kg per day,
while neutralinos typically have event rates in the $10^{-2}$ to
1 event/kg/day range.
Still, even for neutralinos the background is the problem.  There is
a gamma ray background of about 0.5 event/kg/day/keV which seems immune
to shielding, etc.  A very promising recent advance, however, allows
the separation of this gamma ray background from the nuclear recoil
events which make up the signal.  The technique, pioneered by the
Berkeley group,
\refmark\shutt
involves collecting both the ionization electrons
and the phonons.
Since a gamma ray deposits about 33\% of its energy in ionization
and 66\% in phonons,
while a nuclear recoil (e.g. neutron or WIMP) deposits roughly 10\%
in ionization and 90\% in phonons, it is possible to identify and subtract the
gamma ray background.  This allows a factor of up to 100 improvement
in sensitivity.

Figure 3 shows the expected sensitivity of the new Berkeley detector,
as well as the sensitivity of the older detectors, and the expected
event rates for neutralinos in some specific supersymmetric models.
{}From this figure one sees that the new generation of detectors is
capable of probing the neutralino parameter space for the first time.
Thus these small underground detectors are competing with the giant
accelerators in the search for supersymmetry and neutralinos!
And while the areas of parameter space probed overlap somewhat, they
are complementary in other areas.  For example, direct detection techniques can
reach to higher masses than the accelerators.  However, it is also
clear from the figure that this generation of direct detection
experiment is not
capable of performing a definitive experiment.  It is quite possible
for the dark matter to consist of neutralinos, or other WIMPs
and not be seen in any of the current generation experiments.
Still the rate of progress is impressive, and it is exciting that
the favorite dark matter candidates are under attack at last.

Finally, I find it intriguing that the new generation of indirect
detection methods will be probing very similar regions of supersymmetric
parameter space.
Figure 4 shows the area required for an underground proton decay type
detector which would enable it to get a 4 sigma detection of neutrinos
coming from the annihilation of neutralinos in the center of the Sun.
Detectors such as IMB, Frejus, and Kamiokande, lie off the bottom
of this plot, with areas around 400 m$^2$.  The currently running
experiment MACRO has an area of around 1000 m$^2$, which allows it
to just barely probe some regions of neutralino parameter space.
However, it is the newly funded experiments AMANDA (which will take place
in the South Pole ice) and DUMAND (in the Pacific Ocean), with
areas in the $10^6$ m$^2$ range, which will finally be sensitive
to the bulk of SUSY parameter space.
\refmark\indirect
Once again, there is some overlap with accelerators and direct detection
techniques, and some areas in which the indirect searches will be
complementary.

I would like to finish with some comments about the WIMP
models being considered recently.  The neutralino LSP from supersymmetry is by
far the most popular WIMP, but the rates and $\Omega$ depend upon many
parameters.  Of the SUSY models which exist, {\it generic minimal}
models have been investigated the most,
\refmark{\ellishag,\griestneut}
and it was these which were
presented in the previous plots.  These models assume a minimum
of new particles, and only include interactions which preserve $CP$
and R-parity.  They typically use some parameters from Grand Unified
Models (GUTs), but leave others free.  In this way, they are not completely
consistent.  If one considers a GUT, then many parameters are fixed;
however, if one leaves out GUT assumptions, then the parameter space
is less restricted than usually considered.
So, I see two main trends in the theoretical neutralino world.
First, much effort is being made
to work out the minimal Supergravity (SUGRA) model in a consistent
manner.  Here everything is taken from a GUT.  Only 3 or 4 parameters are
needed, and all couplings and masses are derived from these
via the renormalization group.  This allows
more precise predictions, and results in the dark matter neutralino
most probably being a B-ino.
\refmark\minimalsugra
This approach has received recent motivation from the SUSY coupling
constant unification discussed at length by Wilczek (these proceedings).

The other main trend is to move away from the simplest SUGRA model.
There are, of course, many ways to do this.  One can just consider
a more generic minimal model, by relaxing the GUT assumptions,
\refmark{\dreestata,\griestros}
as suggested by superstring models,
\refmark{\ibanez}
or by doing away completely with any
GUT assumptions.
\refmark\randall
One can also consider non-minimal SUSY models
by allowing extra Higgs or gauge structure.
\refmark\extendedsusy
Some of these models have
features which make them quite attractive, but, of course, result in
more free parameters, and predictions which are not as precise or testable.
The important point here, is not to decide which approach is more
realistic or more likely to correspond to the truth, but to warn
the reader that conclusions about SUSY, neutralinos, and neutralino
dark matter, depend sensitively upon the SUSY model being considered,
and that one needs to know which is model is being discussed.

So, I conclude as I began, by noting that for the first time in the
next few years, most of the popular dark matter candidates will
be experimentally accessible, and that with some luck,
we may soon know what the primary constituent of the Universe is.

\bigskip
\centerline{\bf ACKNOWLEDGEMENTS}
I would like to thank R.~Young, M.~Kamionkowski, M.~Spiro, B.~Paczynski,
L.~Roszkowski, and all the members of the MACHO collaboration for advice,
information and help with this work.

\par \penalty-400 \vskip\chapterskip
   \spacecheck\referenceminspace \immediate\closeout\referencewrite
   \referenceopenfalse
   \line{\fourteenrm\hfil REFERENCES\hfil}\vskip\headskip
   \input referenc.texauxil
   
\vfil
\end